\documentclass[conference]{IEEEtran}
\IEEEoverridecommandlockouts
\usepackage{cite}
\usepackage{amsmath,amssymb,amsfonts}
\usepackage{graphicx}
\usepackage{placeins}
\usepackage{textcomp}
\usepackage{xcolor}
\usepackage{float}
\usepackage{stfloats}
\graphicspath{{./images/}{figures/}}
\usepackage[hyphens]{url}
\usepackage[english]{babel}
\usepackage{xspace}
\usepackage{algorithm}
\usepackage{algpseudocode}
\usepackage{pifont}
\usepackage{multirow}
\usepackage{tabularx}
\usepackage{tikz}
\usetikzlibrary{arrows.meta, positioning}
\def\BibTeX{{\rm B\kern-.05em{\sc i\kern-.025em b}\kern-.08em
    T\kern-.1667em\lower.7ex\hbox{E}\kern-.125emX}}

\newcommand{\Sys}{\emph{SciCheck}\xspace}
\newcommand{\FP}{\mathit{FalsePositive}}
\newcommand{\FN}{\mathit{FalseNegative}}

\makeatletter
\newcommand{\linebreakand}{%
  \end{@IEEEauthorhalign}
  \hfill\mbox{}\par
  \mbox{}\hfill\begin{@IEEEauthorhalign}
}
\makeatother

\begin{document}

\title{\Sys: A Framework for Verifying the Integrity of Deep Learning Training}

\author{
\IEEEauthorblockN{Alexander Nguyen}
\IEEEauthorblockA{\textit{Department of Computer Science} \\
\textit{Gonzaga University}\\
alexkhangnguyen@gmail.com}
\and
\IEEEauthorblockN{Dylan Gresham}
\IEEEauthorblockA{\textit{Department of Computer Science} \\
\textit{Boise State University}\\
dylangresham@u.boisestate.edu}
\linebreakand
\IEEEauthorblockN{Gaby G. Dagher}
\IEEEauthorblockA{\textit{Department of Computer Science} \\
\textit{Boise State University}\\
gabydagher@boisestate.edu}
\and
\IEEEauthorblockN{Min Long}
\IEEEauthorblockA{\textit{Department of Computer Science} \\
\textit{Boise State University}\\
minlong@boisestate.edu}
}

\maketitle

\begin{abstract}
Deep learning training is difficult to verify within a scientific workflow: its architecture, hyperparameters, and procedure are rarely fixed until training has been attempted, and two honest runs never produce the same weights, so recomputation cannot serve as a check. We present \Sys, a framework in which the scientist explores training configurations freely and posts a \emph{blueprint} to a provenance ledger for each candidate they want verified, and a verifying scientist judges the resulting run in two layers: a statistical \emph{judge model} that compares per-step gradient statistics against the verifier's own reference run, and a spot-check that re-executes randomly chosen training segments whose checkpoints are committed to the ledger one epoch at a time. On an MLP, a CNN, and an autoencoder at 100k--500k parameters, under gradient sign flipping, label poisoning, and fabricated runs, \Sys reaches F1 of 0.966, 0.906, and 0.976 while rejecting 0--4.4\% of honest runs. For the attacks we test, the spot-check accounts for most of this accuracy, and we characterize what each layer can detect. The effective-signal statistic falls as $(1-2s)^2$ under a sign flip of severity $s$, so gradient statistics alone catch strong enough sign flips, while fabricated runs that replay genuine statistics look no more anomalous than honest runs to any statistical detector we test. The spot-check catches fully fabricated segments with probability at least $1-(1-s)^k$ for $k$ sampled segments when the draws stay hidden until the run is committed, which roughly matches our measurements and lets a verifier choose $k$ for a target soundness. Verification costs 1.5--2.3$\times$ one retraining, mostly for the verifier's reference run; each submission's spot-check takes under 1.5\,s, and per-epoch commitments reduce checkpoint storage 5--17$\times$ compared with keeping every checkpoint.
\end{abstract}

\begin{IEEEkeywords}
Deep Learning; Anomaly Detection; Provenance; Proof-of-Learning; Scientific Workflow;
\end{IEEEkeywords}

\section{Introduction}\label{sec:introduction}

Scientific claims gain credibility through independent repetition, so science must be replicable as well as correct. This self-correcting property breaks down when replication is sparse: later research builds on claims that are eventually invalidated. A seminal 2006 Alzheimer's paper~\cite{lesne2006specific} diverted two decades of resources before most of its authors moved to retract it in 2024 over manipulated images~\cite{piller2024authors}, and psychology's replication crisis left only 36 of 100 influential studies replicable~\cite{pashler2012editors, osc_2015_estimating}. Following the National Academies of Sciences, Engineering, and Medicine~\cite{science2019reproducibility}, we refer to replication and reproduction jointly as repetition.

Machine Learning (ML) faces a version of the same crisis. Deep Learning (DL) has advanced rapidly~\cite{vaswani2017attention, dosovitskiy2021image, santoro2018relational}, but the scale of its experiments makes repeating them costly or infeasible~\cite{science2019reproducibility, jhu2024replication}, and flawed results, such as those caused by data leakage, propagate when later work cites them as trusted~\cite{kapoor_leakage_2023, JMLR:v22:20-303}.

Blockchain-based provenance systems can provide the transparency and tamper-evidence that independent repetition needs~\cite{weidener2024desci}. However, such systems, for example SciLedger~\cite{SciLedger}, assume a task's procedure is fully specified before it executes, an assumption that DL training violates by construction. We propose \Sys, a framework that verifies DL training within a scientific workflow, where one scientist trains and the others act as verifiers, without requiring the training procedure to be pinned down before it has been worked out. The scientist explores freely off the ledger and posts a candidate only when they want it verified; if it is rejected, they return to exploring and post another, and the ledger keeps every attempt.

\Sys verifies a training run in two layers. A statistical judge, built from the verifying scientist's own reference run, checks whether the submitted run's per-step gradient statistics look like honest training; once built, it needs nothing but those statistics. A spot-check re-executes a few randomly chosen segments of the submitted run from checkpoints the scientist committed to the ledger while training. Our evaluation shows what each layer catches: statistics expose attacks that change every gradient step once they are strong enough, but a fabricated run that replays genuine statistics is invisible to them, and only re-execution catches it, with a probability the verifier can set in advance.

\subsection{Contributions}\label{sec:contributions}

\begin{itemize}
    \setlength\itemsep{0em}
    \item We propose \Sys, a framework that verifies DL training within a scientific workflow. The scientist explores off the ledger and posts a blueprint for each candidate they want verified, and the resulting run is judged in two layers: a statistical judge over per-step training snapshots, and a checkpoint spot-check whose commitments are recorded on a SciLedger-style ledger one epoch at a time, which bounds the scientist's storage to one epoch.
    \item We explain what each layer can and cannot detect. We derive that the effective-signal statistic scales as $(1-2s)^2$ under gradient sign flipping, which fixes the attack severity the effective-signal check catches; show that none of the statistical detectors we test separates fabricated runs from honest ones; and derive the spot-check's soundness $1-(1-s)^k$. All three agree with our measurements.
    \item We evaluate \Sys on the full adversary model (gradient tampering, label poisoning, and fabrication) across an MLP, a CNN, and an autoencoder at 100k--500k parameters, with 2,700 scientist runs. With per-architecture calibration under a 5\% false-positive limit, its statistical judge outperforms Isolation Forest and One-Class SVM baselines calibrated the same way, mostly through the effective-signal check.
\end{itemize}

\section{Related Work}\label{sec:related_work}

\subsection{Verifying Scientific Data and Process Integrity}

Provenance is essential to reproducible scientific workflows~\cite{davidson_provenance_2008}, and a range of systems make it tamper-evident. SciBlock~\cite{sciblock} provides non-repudiable storage for workflow provenance, SciChain~\cite{al-mamun_scichain_2021} records provenance for high-performance computing with low overhead, and VAULT~\cite{gazsi_vault_2021} supports secure multi-party data access through quorum-based consensus. Work on smart-contract accountability~\cite{neisse2017}, provenance for AI assets~\cite{luthi_distributed_2020}, and certified research processes~\cite{wittek_blockchain-based_2021} further improves auditability~\cite{tosh2017consensus, lu_integration_2024}. SciLedger~\cite{SciLedger} is representative of a limitation these systems share: its provenance guarantee rests on every task being fully specified before it executes, which DL training violates by construction (\S\ref{sec:problem_formulation}). \Sys closes this gap with a fixed-blueprint training task, a \emph{judge model}, and per-epoch checkpoint commitments that a SciLedger-style system can adopt. PRISM~\cite{miller_prism_2023} is complementary: its reputation-based consensus strengthens \emph{who} validates a task, not \emph{what} it means to validate a nondeterministic one.

\subsection{Verifying Machine Learning Data and Processes}

Most tools for ML integrity validate the data a model is trained on rather than the training process~\cite{kapoor_leakage_2023, gardner_enabling_2018, barbierato_challenges_2024, Data_Lifecycle_Challenges, shankar_automatic_2023, chai_data_2023}, and none produce an independently auditable record. Work that does check computation, such as performance predictors for faulty inputs~\cite{black_box_predictions} and verifiable execution~\cite{SafetyNets}, assumes an external, incentivized validator that scientific research lacks.

Proof-of-Learning (PoL)~\cite{jia2021proof} verifies that a model was trained as claimed: the trainer logs intermediate checkpoints and the data used between them, and a verifier re-executes part of the log and compares the result with the logged checkpoints. Later work shows that an adversary can construct logs that pass this check at lower cost than honest training~\cite{zhang2022adversarial, fang2023proof}. \Sys's spot-check follows PoL's re-execution idea but adapts it to a scientific workflow: the scientist commits each epoch's checkpoints to the provenance ledger as a single Merkle root~\cite{merkle1987digital} and keeps only the segments later drawn for checking, so storage stays bounded to one epoch, and a statistical judge that needs no checkpoints at all screens every submission first.

Closest to \Sys's judge model is FIDELIS~\cite{carney_fidelis_2025}, which detects poisoned updates in federated learning through consensus among client-produced judge models; \Sys's judge instead belongs to a single verifying scientist and is applied to an entire submitted run after the fact. Finally, work on neural architecture search and automated ML treats a model's configuration and training procedure as an open search problem~\cite{elsken_neural_2019, he_automl_2021, zoller_benchmark_2021, shen_automated_2024}, which is why a provenance system cannot demand that a training procedure be fixed up front.

\section{Problem Formulation}\label{sec:problem_formulation}

\subsection{Problem Setting}

A normal workflow task on SciLedger locks in its procedure before it runs, so a verifier can later check the recorded output against that procedure. A training task cannot: the scientist typically does not know the final architecture, hyperparameters, or procedure until training has been attempted, and committing one in advance would force a choice between committing before knowing whether it works and leaving the specification too vague to guarantee anything. The workflow can still fix some things up front, such as the dataset and what the finished model must achieve. We call this partial commitment, the fixed constraints plus whatever the scientist still has to decide, a specification, written $\mathsf{Spec}$. \Sys therefore never commits the training procedure when the task is created; it commits one candidate procedure at a time, after the scientist has explored (\S\ref{sec:solution_overview}).

\subsection{Adversary Model}

We assume a malicious scientist submitting a training run under some specification $\mathsf{Spec}$ could:
\begin{itemize}
    \setlength\itemsep{0em}
    \item Train on data they are not supposed to use, breaking one of $\mathsf{Spec}$'s fixed constraints;
    \item Tamper with an otherwise normal training run, for example by flipping the sign of gradients;
    \item Skip training altogether and submit a fabricated result.
\end{itemize}
In every case, the goal is to have a run accepted even though it violates $\mathsf{Spec}$. Our experiments instantiate these adversaries as label poisoning, gradient sign flipping, and fabricated runs (\S\ref{sec:acc_setup}). We assume the verifying scientist is semi-honest; a dishonest or colluding verifier is left to future work on reputation and consensus mechanisms.

\subsection{Limitations of Direct Recomputation}\label{sec:recompute}

For an ordinary task, checking validity means redoing the computation and comparing. That does not work for training: two runs from different random initializations, under exactly the same procedure, land on different and equally valid parameters, so there is no single correct result to recompute against. Validity must instead be reframed as ``does this look like something a genuine run of this procedure would produce.'' Short stretches of a run are different: from a committed starting state and a committed sequence of batches, a few optimizer steps are reproducible up to floating-point nondeterminism. \Sys uses both views.

\subsection{Correctness Goal}

Table~\ref{tbl:ProblemNotation} summarizes our notation. Let $\mathcal{H}_{\mathsf{Spec}}$ be the range of training records an honest scientist would produce by following $\mathsf{Spec}$. Given a submitted record $T$, we want a check $\mathsf{Validate}(\mathsf{Spec}, T) \in \{\mathsf{accept}, \mathsf{reject}\}$ with two properties:

\textbf{Completeness: honest work should not be rejected.} For $T \in \mathcal{H}_{\mathsf{Spec}}$,
\[
\Pr[\mathsf{Validate}(\mathsf{Spec}, T) = \mathsf{accept}] \ge 1 - \FP.
\]

\textbf{Soundness: dishonest work should not be accepted.} For any $T$ the attacker produces that violates $\mathsf{Spec}$,
\[
\Pr[\mathsf{Validate}(\mathsf{Spec}, T) = \mathsf{accept}] \le \FN.
\]

$\FP$ and $\FN$ pull against each other: a stricter check catches more attackers but rejects more honest work. For \Sys's statistical layer these are empirically calibrated guarantees, not cryptographic ones. The spot-check layer adds a probabilistic soundness bound against fabrication, which we derive in \S\ref{sec:analysis_spot}. Neither layer is a formal proof system.

\begin{table*}[!ht]
    \centering
    \small
    \renewcommand{\arraystretch}{1.15}
    \caption{List of Notations}
    \label{tbl:ProblemNotation}
    \begin{tabularx}{\linewidth}{ |r|X|r|X| }
        \hline
        Symbol & Description & Symbol & Description \\
        \hline
        $\FP$ & Chance an honest run gets wrongly rejected & $\theta$ & Acceptance threshold on $\phi$ \\
        $\FN$ & Chance a cheating run gets wrongly accepted & $\eta_t$ & Effective-signal statistic of snapshot $s_t$, $\eta_t = \overline{g}_t^{2}/(\overline{g}_t^{2}+\sigma_t^{2})$ \\
        $\hat b$ & Blueprint: one candidate's dataset, architecture, hyperparameters, and procedure, posted before training & $\bar\eta_v$ & Mean effective signal over the verifying scientist's training snapshots \\
        $s_t$, $T$ & Snapshot at batch $t$; snapshot sequence & $\rho$, $\gamma$ & Effective-signal ratio and acceptance cutoff $\gamma = \rho \cdot \bar\eta_v$ \\
        $J$, $W$ & Judge model (Anomaly Transformer) and its window length & $\psi$, $\psi_{\max}$ & Share of steps repeating the step one epoch earlier, and its limit \\
        $\tau$ & Per-timestep anomaly-score threshold & $k$, $\ell$ & Number and length (in steps) of spot-checked segments \\
        $\phi$ & Fraction of timesteps scored above $\tau$ & $\delta$, $R_e$ & Spot-check tolerance; Merkle root of epoch $e$'s checkpoints \\
        \hline
    \end{tabularx}
\end{table*}

\section{Solution Overview}\label{sec:solution_overview}

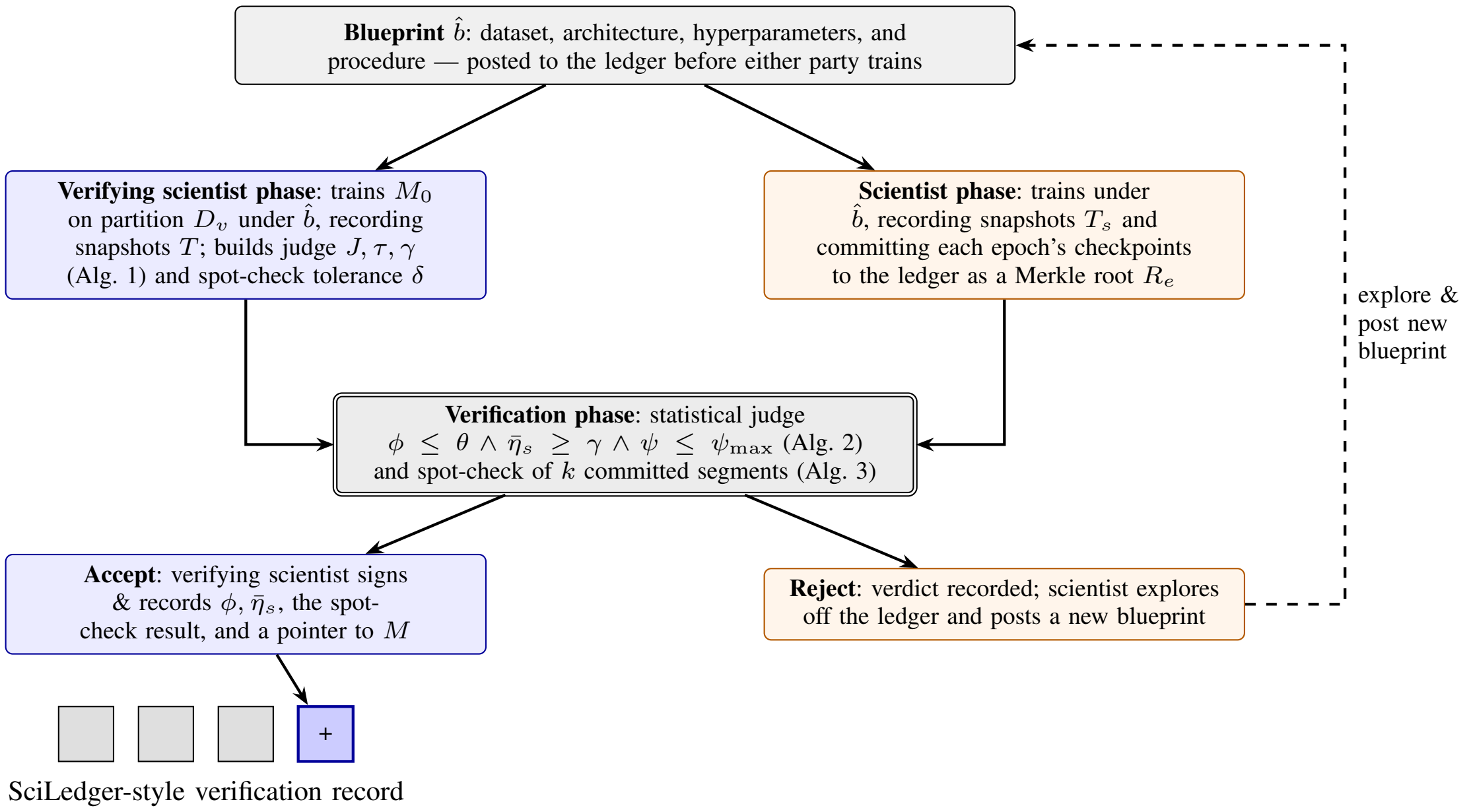
\begin{figure*}[!t]
\centering
\begin{tikzpicture}[
  every node/.style={font=\scriptsize},
  bpnode/.style={rectangle, draw, rounded corners=2pt, fill=gray!12, text width=7.6cm, align=center, inner sep=3pt},
  mnode/.style={rectangle, draw=blue!60!black, rounded corners=2pt, fill=blue!8, text width=4.6cm, align=center, inner sep=3pt},
  snode/.style={rectangle, draw=orange!70!black, rounded corners=2pt, fill=orange!8, text width=4.6cm, align=center, inner sep=3pt},
  dec/.style={rectangle, draw, double, rounded corners=2pt, fill=gray!15, text width=5.6cm, align=center, inner sep=3pt},
  blk/.style={rectangle, draw, minimum width=0.55cm, minimum height=0.55cm, fill=gray!25},
  newblk/.style={blk, draw=blue!60!black, thick, fill=blue!20},
  arr/.style={-{Stealth[length=1.8mm]}, thick}
]

\node (bp) [bpnode] at (0,0) {\textbf{Blueprint $\hat b$}: dataset, architecture, hyperparameters, and procedure --- posted to the ledger before either party trains};

\node (miner) [mnode] at (-3.8,-1.9) {\textbf{Verifying scientist phase}: trains $M_0$ on partition $D_v$ under $\hat b$, recording snapshots $T$; builds judge $J,\tau,\gamma$ (Alg.~\ref{alg:judge}) and spot-check tolerance $\delta$};
\node (sci)   [snode] at (3.8,-1.9) {\textbf{Scientist phase}: trains under $\hat b$, recording snapshots $T_s$ and committing each epoch's checkpoints to the ledger as a Merkle root $R_e$};

\node (val) [dec] at (0,-4.0) {\textbf{Verification phase}: statistical judge $\phi \le \theta \wedge \bar\eta_s \ge \gamma \wedge \psi \le \psi_{\max}$ (Alg.~\ref{alg:validate}) and spot-check of $k$ committed segments (Alg.~\ref{alg:spot})};

\node (acc) [mnode] at (-3.8,-5.6) {\textbf{Accept}: verifying scientist signs \& records $\phi,\bar\eta_s$, the spot-check result, and a pointer to $M$};
\node (rej) [snode] at (3.8,-5.6) {\textbf{Reject}: verdict recorded; scientist explores off the ledger and posts a new blueprint};

\node (c1) [blk] at (-5.4,-6.9) {};
\node (c2) [blk] at (-4.6,-6.9) {};
\node (c3) [blk] at (-3.8,-6.9) {};
\node (c4) [newblk] at (-3.0,-6.9) {+};
\node at (-4.2,-7.5) {\footnotesize SciLedger-style verification record};

\draw[arr] (bp) -- (miner);
\draw[arr] (bp) -- (sci);
\draw[arr] (miner.south) |- (val.west);
\draw[arr] (sci.south) |- (val.east);
\draw[arr] (val) -- (acc);
\draw[arr] (val) -- (rej);
\draw[arr] (acc) -- (c4);
\draw[arr, dashed] (rej.east) -- ++(1.0,0) |- node[pos=0.25, right, align=left, font=\scriptsize] {explore \&\\post new\\blueprint} (bp.east);

\end{tikzpicture}
\caption{Overview of the \Sys pipeline. Both parties train under a shared blueprint. The verifying scientist builds a judge model and a spot-check tolerance from their own run; the scientist commits checkpoints to the ledger while training. Every blueprint is posted to the ledger before training. An accepted run is signed into a SciLedger-style verification record; after a rejection, the scientist returns to exploring off the ledger and posts a new blueprint, so the ledger keeps the history of every attempt.}
\label{fig:overview}
\end{figure*}

\Sys closes this gap by separating exploration from verification. The scientist explores architectures, hyperparameters, and procedures freely off the ledger, within $\mathsf{Spec}$'s fixed constraints, and none of that exploration needs to be recorded. When a candidate is ready, the scientist posts a \emph{blueprint} to a SciLedger-style tamper-evident ledger~\cite{SciLedger}: a full specification of that one candidate training run's dataset, architecture, hyperparameters, and procedure, fixed before either party trains under it. Two parties then train independently under the blueprint: the scientist, and a verifying scientist assigned to validate the scientist's work (Figure~\ref{fig:overview}). If the run is rejected, or the trained model falls short of $\mathsf{Spec}$'s goals, the scientist returns to exploring and posts a new blueprint. A blueprint therefore fixes one attempt, not the whole task: the ledger keeps every posted blueprint with its verdict, so it records the history of what was tried without requiring the final procedure up front. \Sys verifies that a posted run was trained as its blueprint states; it does not verify how the scientist arrived at that blueprint.

The verifying scientist trains a reference model $M_0$ on their own data partition, recording a \emph{snapshot} $s_t$ of gradient and weight statistics once per batch (\S\ref{sec:snapshot_representation}). From these snapshots they build a \emph{judge model} $J$, an Anomaly Transformer (AT)~\cite{xu2022anomalytransformer}, together with an effective-signal baseline $\bar\eta_v$ (Algorithm~\ref{alg:judge}), and they re-execute segments of their own run to calibrate the spot-check tolerance $\delta$ (\S\ref{sec:spot_check}). The scientist trains under the same blueprint, records snapshots in the same format, and commits each epoch's checkpoints (weights, optimizer state, batch order, and augmentation state) to the ledger as a single Merkle root. Verification applies the statistical judge to the submitted snapshots $T_s$ (Algorithm~\ref{alg:validate}) and re-executes $k$ randomly drawn committed segments (Algorithm~\ref{alg:spot}). A run that passes every check is signed and recorded with a pointer to the trained model; a rejected run is recorded with its verdict, and the scientist returns to exploring.

\section{Solution: \Sys Framework}\label{sec:solution}

\subsection{Snapshot Representation}\label{sec:snapshot_representation}

Each snapshot $s_t$ captures $13 + L$ statistics computed once per batch from the gradient $g_t$ and the weights $w_t$ across the $L$ tracked layers (Table~\ref{tbl:SnapshotFeatures}): seven training-dynamics statistics chosen to expose Byzantine gradient manipulation~\cite{blanchard2017byzantine} such as sign flipping~\cite{li2019rsa}, six describing the shape of the gradient distribution, and one gradient norm per tracked layer. Every entry is a scalar aggregate rather than a raw parameter value, which makes snapshots comparable across differently initialized models. Snapshots are taken from the gradient the optimizer actually applies, after any tampering.

\begin{table}[!t]
    \centering
    \footnotesize
    \renewcommand{\arraystretch}{1.1}
    \caption{The $13 + L$ statistics in each snapshot $s_t$, where $\bar g_t$ and $\sigma_t$ are the mean and standard deviation of $g_t$ over all parameters and $L$ is the number of tracked layers.}
    \label{tbl:SnapshotFeatures}
    \begin{tabularx}{\linewidth}{ |l|l|X| }
        \hline
        Statistic & Definition & What it exposes \\
        \hline
        Signal-noise ratio & $|\bar g_t| / \sigma_t$ & Driven toward zero by sign flips \\
        Loss delta & $\text{loss}_t - \text{loss}_{t-1}$ & Erratic swings in corrupted runs \\
        Effective signal $\eta_t$ & $\bar g_t^2 / (\bar g_t^2 + \sigma_t^2)$ & Share of consistent signal; basis of the $\gamma$ check \\
        Gradient norm & $\lVert g_t \rVert_2$ & Scaled-up or suppressed gradients \\
        Weight-delta norm & $\lVert w_t - w_{t-1} \rVert_2$ & Direct weight tampering \\
        Raw loss & $\text{loss}_t$ & Position along the trajectory \\
        Weight norm & $\lVert w_t \rVert_2$ & Divergence or fabricated weights \\
        Fraction positive & $\Pr[g_{t,i} > 0]$ & Pushed to an even split by sign flips \\
        Percentiles & $q_{10}, q_{50}, q_{90}$ of $|g_{t,i}|$ & Spread of gradient magnitudes \\
        Skewness & $\mathbb{E}[z^3]$ of standardized $g_t$ & Asymmetry of the gradients \\
        Kurtosis & $\mathbb{E}[z^4] - 3$ & A few extreme outlier gradients \\
        Layer norms & $\lVert g_t^{(\ell)} \rVert_2$ per layer $\ell$ & Anomalies confined to one layer \\
        \hline
    \end{tabularx}
\end{table}

\begin{algorithm}[!t]
    \small
    \caption{Training of Judge Model $J$}
    \label{alg:judge}
    \begin{algorithmic}[1]
        \Require Blueprint $\hat b$ (fixes model $M_0$), verifying scientist's data partition $D_v$, window length $W$, association-discrepancy weight $\lambda$, threshold percentile $r$, effective-signal ratio $\rho$
        \Ensure Judge model $J$, anomaly threshold $\tau$, effective-signal cutoff $\gamma$
        \State Train $M_0$ on $D_v$ per $\hat b$, recording one snapshot $s_t$ per batch, giving $T = (s_1, \dots, s_N)$
        \State Split $T$ chronologically into $T_{train}$, $T_{val}$, $T_{test}$ and normalize all three with the mean and standard deviation of $T_{train}$
        \State Train $J$ on windows of length $W$ over $T_{train}$, minimizing reconstruction error while maximizing the prior/series association discrepancy, weighted by $\lambda$
        \State Score the windows of $T_{val}$ with $J$ and reassemble the scores into a per-timestep anomaly timeline
        \State $\tau \gets$ the $(1-r)$-quantile of that timeline
        \State $\bar\eta_v \gets$ the mean of $\eta_t$ over $T_{train}$; \quad $\gamma \gets \rho \cdot \bar\eta_v$
        \State \Return $J, \tau, \gamma$
    \end{algorithmic}
\end{algorithm}

\subsection{Judge Model}\label{sec:at_backbone}

The judge model $J$ is the association-discrepancy Anomaly Transformer of Xu \textit{et al.}~\cite{xu2022anomalytransformer}. At each layer $\ell$, $J$ contrasts a learned \emph{series-association} $S^{(\ell)}$ (self-attention over the window) with a fixed \emph{prior-association} $P^{(\ell)}$ (a Gaussian kernel favoring temporal neighbors), and computes their discrepancy, averaged over layers:
\begin{equation}
\label{eq:ass_dis}
\mathrm{AssDis}(P^{(\ell)}, S^{(\ell)}) = \mathrm{KL}(P^{(\ell)} \Vert S^{(\ell)}) + \mathrm{KL}(S^{(\ell)} \Vert P^{(\ell)}).
\end{equation}
$J$ is trained with Xu \textit{et al.}'s minimax strategy, so the discrepancy stays small for snapshots that resemble their neighbors and large for those that do not. Each window's anomaly score combines this discrepancy with reconstruction error~\cite{xu2022anomalytransformer}.

\subsection{Validating a Submitted Run}\label{sec:validating_run}

The statistical checks are computed once, after the scientist's run has finished, since $\phi$, $\bar\eta_s$, and $\psi$ are aggregates over the whole run (Algorithm~\ref{alg:validate}).

\paragraph{Replay check} A forger who skips part of training could fill the missing record by copying an earlier stretch of genuine snapshots. Honest training never repeats itself, so the replay check compares each normalized snapshot with the one exactly an epoch earlier (the blueprint fixes the epoch length $E$) and computes $\psi$, the share of steps whose distance to it is at most $\kappa$ times the run's median step-to-step distance. The run must satisfy $\psi \le \psi_{\max}$. This catches only exact or near-exact replays; a forger who adds noise of the size of natural step-to-step variation evades it, which is why \Sys also spot-checks committed checkpoints.

\begin{algorithm}[!t]
\small
\caption{Statistical Judge: Validating a Submitted Run}
\label{alg:validate}
\begin{algorithmic}[1]
\Require $J$, $\tau$, $\gamma$, $\theta$, $W$, the verifying scientist's normalization statistics $(\mu_v, \sigma_v)$, epoch length $E$, replay tolerance $\kappa$, replay limit $\psi_{\max}$, submitted snapshots $T_s = (s_1, \dots, s_M)$
\Ensure Decision $\in \{\mathsf{accept}, \mathsf{reject}\}$
\State $\tilde T_s \gets T_s$ normalized with $(\mu_v, \sigma_v)$
\State Score windows of length $W$ over $\tilde T_s$ with $J$ and reassemble them into a per-timestep anomaly timeline
\State $\phi \gets$ the fraction of timesteps in the timeline exceeding $\tau$
\State $\bar\eta_s \gets$ the mean of $\eta_t$ over all snapshots in $T_s$
\State $\psi \gets$ the fraction of $t$ with $\lVert \tilde s_{t+E} - \tilde s_t \rVert \le \kappa \cdot \mathrm{median}_t \lVert \tilde s_{t+1} - \tilde s_t \rVert$
\State \Return $\mathsf{accept}$ if $\phi \le \theta$, $\bar\eta_s \ge \gamma$, and $\psi \le \psi_{\max}$; otherwise $\mathsf{reject}$
\end{algorithmic}
\end{algorithm}

\subsection{Committed Checkpoints and the Spot-Check}\label{sec:spot_check}

Statistics alone cannot show that the submitted run was actually computed (\S\ref{sec:analysis_fabrication}). \Sys therefore also checks a sample of the computation itself, following the re-execution idea of Proof-of-Learning~\cite{jia2021proof}.

\paragraph{Commitments} The scientist's run is divided into segments of $\ell$ optimizer steps. For each segment, the scientist records a checkpoint at its start (weights, optimizer state, and augmentation state), the indices of the batches it consumes, and the weights at its end. At the end of each epoch $e$, the scientist hashes that epoch's checkpoints into a Merkle tree~\cite{merkle1987digital} and records its root $R_e$ on the ledger. The verifier checks $k$ of the run's $N/\ell$ segments, drawn uniformly at random from randomness that only exists after the commitment, such as the hash of the next block, so the scientist cannot know in advance which segments will be checked. The scientist then keeps the drawn segments and deletes the rest, so at most one epoch of checkpoints, plus the segments drawn so far, is on disk at any time. Revealing a fixed total of $k$ draws one epoch at a time has a weakness we discuss in \S\ref{sec:limitations}.

\paragraph{Re-execution} For each of the $k$ drawn segments, the verifying scientist checks the segment's data against $R_e$ with its Merkle path, restores the starting checkpoint, re-runs $\ell$ steps on the blueprint's data with the committed batches, and measures how far the result ends from the committed end weights, relative to how far the segment moved (Algorithm~\ref{alg:spot}). An honest segment reproduces up to floating-point nondeterminism; a segment whose gradients were tampered with, whose data were not the blueprint's, or that was never computed does not.

\paragraph{Tolerance} GPU kernels are not always bit-reproducible, so the verifying scientist calibrates the tolerance on their own run: they re-execute segments of $M_0$ the same way and set $\delta = \max(m \cdot \max_i e_i^{(v)}, \delta_{\min})$, a margin $m$ above the largest error their own honest segments show.

\begin{algorithm}[!t]
\small
\caption{Spot-Check of Committed Segments}
\label{alg:spot}
\begin{algorithmic}[1]
\Require Ledger roots $R_e$, segment length $\ell$, number of segments $k$, tolerance $\delta$, blueprint data $D$, randomness revealed after each epoch's commitment
\Ensure Decision $\in \{\mathsf{accept}, \mathsf{reject}\}$
\State Draw $k$ of the $N/\ell$ segments uniformly without replacement, using the post-commitment randomness
\For{each drawn segment $i$}
    \State Receive start checkpoint $c_i$ (weights $w_i$, optimizer and augmentation state), batch indices $B_i$, end weights $w_i'$, and Merkle paths; verify them against $R_e$
    \State $\hat w_i \gets$ run $\ell$ optimizer steps from $c_i$ on $D$ with batches $B_i$
    \State $e_i \gets \lVert \hat w_i - w_i' \rVert / \lVert \hat w_i - w_i \rVert$
\EndFor
\State \Return $\mathsf{accept}$ if $\max_i e_i \le \delta$; otherwise $\mathsf{reject}$
\end{algorithmic}
\end{algorithm}

\subsection{Formal Acceptance Predicate}\label{sec:accept_predicate}

Algorithms~\ref{alg:validate} and~\ref{alg:spot} together are \Sys's concrete instantiation of the abstract $\mathsf{Validate}$ check from \S\ref{sec:problem_formulation}:
\[
\mathsf{Validate}(\mathsf{Spec}, T_s) =
\begin{cases}
\mathsf{accept} & \text{if } \phi \le \theta \wedge \bar\eta_s \ge \gamma \\
 & \quad \wedge\, \psi \le \psi_{\max} \wedge \max_i e_i \le \delta, \\
\mathsf{reject} & \text{otherwise.}
\end{cases}
\]
The two layers play different roles. The statistical judge needs only the snapshot sequence, costs one pass of the judge over it, and targets attacks that change the gradient statistics of every step. The spot-check needs committed checkpoints and re-execution, and targets what statistics cannot see: a run, or part of a run, that was never computed. The snapshots are reported by the scientist, so the statistical judge only catches a scientist who reports their true statistics; one who submits statistics copied from a separate honest run passes it. For the attacks we test, soundness therefore comes from the spot-check, which re-executes the committed computation; the statistical judge is a screen that works without checkpoints, though building it requires the verifier's full reference run. \S\ref{sec:analysis} derives what each layer catches. Full integration into a concrete provenance system is left to future work; our experiments evaluate both layers independently of any ledger deployment.

\section{Experimental Evaluation\protect\footnote{Claude assisted with portions of the experimental codebase and with organizing, processing, and analyzing the raw experimental outputs reported in this section. The authors reviewed and verified all resulting code, analyses, and reported results. See the Disclosure on the Use of AI Tools at the end of this paper for further details.}}\label{sec:experimental_evaluation}

\subsection{Setup}\label{sec:acc_setup}

\paragraph{Architectures and data} We instantiate the protocol three times, once per architecture (Table~\ref{tbl:AccSetup}). A Multi-Layer Perceptron (MLP) with two hidden layers is trained for binary signal-versus-background classification on the first 200,000 rows of the HIGGS dataset~\cite{baldi2014searching} (28 standardized features). A Convolutional Neural Network (CNN) with two convolutional layers and a linear head is trained for cat-versus-dog classification on all of PetImages~\cite{elson2007asirra} (24,998 images resized to $32\times32$ RGB, with random flip and crop augmentation). An Autoencoder (AE) with one hidden layer is trained for reconstruction on all 60,000 MNIST training digits~\cite{lecun1998gradient}. In every case 15\% of the data is held out from both parties; every scientist, honest or malicious, trains on the remaining 85\%, and the verifying scientist trains on the first 30\% of it.

\begin{table}[!t]
\centering
\small
\renewcommand{\arraystretch}{1.1}
\caption{Rows each party trains on and rows held out from both. Snapshot dim.\ is $13+L$, where $L$ is the number of tracked layers.}
\label{tbl:AccSetup}
\begin{tabularx}{\linewidth}{|l|X|c|c|c|c|}
\hline
Arch. & Dataset & Ver. Sci. & Sci. & Held out & Dim. \\
\hline
MLP & HIGGS & 51{,}000 & 170{,}000 & 30{,}000 & 16 \\
CNN & PetImages & 6{,}374 & 21{,}248 & 3{,}750 & 16 \\
AE & MNIST & 15{,}300 & 51{,}000 & 9{,}000 & 15 \\
\hline
\end{tabularx}
\end{table}

\paragraph{Models and training} Each architecture is trained at about 100k, 300k, and 500k parameters (99,631--501,321) by varying one width parameter. All models train for 25 epochs with Adam~\cite{kingma2015adam} (learning rate $10^{-3}$, batch size 128). The reference model $M_0$ trains for as many optimizer steps as a scientist takes over the full data, i.e.\ about 3.3 times as many epochs over its smaller partition.

\paragraph{Attacks} Each attack runs at severities $s \in \{0.1, 0.3, 0.5\}$:
\begin{itemize}
    \setlength\itemsep{0em}
    \item \textbf{Sign flip}: for every parameter tensor, an independent Bernoulli($s$) mask is redrawn every batch and the masked gradient entries are sign-flipped before the optimizer step.
    \item \textbf{Label poisoning}: a fraction $s$ of the training rows, drawn once, is corrupted: labels are flipped for the classifiers and pixels inverted for the autoencoder.
    \item \textbf{Fabricated run}: the scientist trains only the first $1-s$ of the steps and fills the rest of the record by replaying their last genuine epoch of snapshots with Gaussian noise of 0.1 times each statistic's step-to-step standard deviation. For the spot-check, they commit their last genuine checkpoint for every skipped step.
\end{itemize}

\paragraph{Checks and baselines} $J$ has $d_{\text{model}}=64$, 4 heads, $d_{ff}=128$, 2 encoder layers, $\lambda=3.0$, and $W=30$, and trains for up to 15 epochs (patience 5, learning rate $10^{-4}$, batch size 32). The reference snapshots are split 70/15/15, with $r=0.01$, so $\tau$ is the 99th percentile of $J$'s scores on the verifier's own validation split. The replay check uses $\kappa = 0.02$ and $\psi_{\max}=0.05$; the spot-check draws $k=10$ segments of $\ell=50$ steps with $m=10$ and $\delta_{\min}=10^{-4}$. Unless stated otherwise, $\theta=0.3$ and $\rho=0.4$. As baselines we fit an Isolation Forest~\cite{liu2008isolation} (200 trees) and a One-Class SVM~\cite{scholkopf2001estimating} (RBF kernel, $\nu = 0.05$) on the same reference snapshots; each sets its own $\tau$ at the same quantile and flags a run by its own $\phi$, alone or combined with the $\eta$ check.

\paragraph{Runs and metrics} Each (architecture, size, seed) unit has one verifying scientist and one judge, which checks 15 honest runs and 5 malicious runs per (attack, severity) cell. With 3 architectures, 3 sizes, and 3 seeds, the main experiment trains $27 \times 60 = 1{,}620$ scientist runs. Two controls re-train the 300k scientist runs with the same seeds and change only the verifier, adding 1,080 runs. We report F1 per cell (a cell's 5 malicious runs against the unit's 15 honest runs), averaged over cells, with a rejected malicious run as a true positive, and $\FP$ and $\FN$ from pooled counts. Because every cell has 5 malicious and 15 honest runs, rejecting every run scores F1 $=0.40$, which we use as a floor.

\subsection{Detection Accuracy}\label{sec:acc_results}

Table~\ref{tbl:Accuracy} reports the full decision. \Sys reaches F1 of 0.966 (MLP), 0.906 (CNN), and 0.976 (AE), rejecting no honest MLP or AE runs and 4.4\% of honest CNN runs, all at 300k and 500k. Every sign-flipped and every label-poisoned run is rejected, at every severity and size. The only misses are fabricated runs: 31--47\% are accepted at $s=0.1$, 2--4\% at $s=0.3$, and none at $s=0.5$.

\begin{table*}[!t]
\centering
\small
\renewcommand{\arraystretch}{1.1}
\caption{Full decision (statistical judge and spot-check) on the main experiment at $\theta=0.3$ and $\rho=0.4$. F1 is averaged over cells; $\FP$ and $\FN$ pool raw counts. Sign flip and label poisoning give identical results at every severity.}
\label{tbl:Accuracy}
\resizebox{\ifdim\width>\linewidth\linewidth\else\width\fi}{!}{%
\begin{tabular}{|l|c|c|c|c|c|c|c|c|c|}
\hline
\multirow{2}{*}{} & \multicolumn{3}{c|}{MLP} & \multicolumn{3}{c|}{CNN} & \multicolumn{3}{c|}{AE} \\
\cline{2-10}
 & F1 & $\FP$ & $\FN$ & F1 & $\FP$ & $\FN$ & F1 & $\FP$ & $\FN$ \\
\hline
100k & 0.955 & 0.000 & 0.052 & 0.959 & 0.000 & 0.059 & 0.976 & 0.000 & 0.037 \\
300k & 0.976 & 0.000 & 0.037 & 0.887 & 0.067 & 0.044 & 0.983 & 0.000 & 0.030 \\
500k & 0.967 & 0.000 & 0.044 & 0.873 & 0.067 & 0.067 & 0.968 & 0.000 & 0.044 \\
\hline
Sign flip (any $s$) & 1.000 & 0.000 & 0.000 & 0.943 & 0.044 & 0.000 & 1.000 & 0.000 & 0.000 \\
Label poisoning (any $s$) & 1.000 & 0.000 & 0.000 & 0.943 & 0.044 & 0.000 & 1.000 & 0.000 & 0.000 \\
Fabricated run, $s=0.1$ & 0.705 & 0.000 & 0.378 & 0.639 & 0.044 & 0.467 & 0.792 & 0.000 & 0.311 \\
Fabricated run, $s=0.3$ & 0.988 & 0.000 & 0.022 & 0.918 & 0.044 & 0.044 & 0.988 & 0.000 & 0.022 \\
Fabricated run, $s=0.5$ & 1.000 & 0.000 & 0.000 & 0.943 & 0.044 & 0.000 & 1.000 & 0.000 & 0.000 \\
\hline
\textbf{Overall} & 0.966 & 0.000 & 0.044 & 0.906 & 0.044 & 0.057 & 0.976 & 0.000 & 0.037 \\
\hline
\end{tabular}%
}
\end{table*}

Applying each check on its own shows which one does the work. The spot-check alone matches the full decision on the MLP and AE and slightly exceeds it on the CNN (F1 0.922), where the $\eta$ check adds false positives. At $\theta=0.3$ the AT rejects almost no runs (F1 0.000, 0.000, and 0.012), so the statistical judge decides exactly as the $\eta$ check does alone, and the replay check never fires, because the fabricated runs' noise keeps $\psi$ near zero. The verifier's own training affects at most how many honest runs are rejected. In the two 300k controls, giving the verifier all of the scientist's data instead of 30\% makes no difference beyond seed-to-seed variation. The main experiment's three CNN false positives at 300k all come from the spot-check, whose tolerance is calibrated on the verifier's own run; both controls train a new verifier run, and neither has any spot-check false positives, including the step-matching control, where the verifier still trains on 30\%. Training the verifier for as many epochs as the scientist, instead of as many steps, makes the $\eta$ check reject honest runs instead: 8.9\% on the MLP (from 0) and 4.4\% on the CNN. In both controls the share of attacks accepted is unchanged.

\subsection{The Statistical Judge Against the Baselines}\label{sec:calibration}

When checkpoints are unavailable, the statistical judge must stand on its own, and a fixed $\theta$ and $\rho$ for every architecture and detector is not a fair comparison: at the original settings the One-Class SVM scores higher than the judge on the MLP and AE, but only by rejecting 24\% and 13\% of honest runs (34\% on the CNN). The judge's own F1 there (0.278 and 0.315) is below the reject-all floor of 0.40, which F1 alone makes look worse than it is: rejecting every run also rejects every honest one, while the judge rejects at most 1.5\%. We therefore calibrate every detector the same way: a grid search over $\theta$ (0 to 1) and $\rho$ (0 to 1.5) per architecture, choosing the highest mean F1 that rejects fewer than 5\% of honest runs for every seed's judge, selected on two seeds and scored on the third. All decisions are recomputed from the stored statistics, without retraining.

\begin{table}[!t]
\centering
\small
\renewcommand{\arraystretch}{1.1}
\setlength{\tabcolsep}{3pt}
\caption{F1 of snapshot-only detectors at the original settings ($\theta=0.3$, $\rho=0.4$) and calibrated per architecture with fewer than 5\% of honest runs rejected per seed. Calibrated F1 is measured on a seed not used for calibration.}
\label{tbl:Calibration}
\resizebox{\ifdim\width>\linewidth\linewidth\else\width\fi}{!}{%
\begin{tabular}{|l|c|c|c|c|c|c|}
\hline
\multirow{2}{*}{Detector} & \multicolumn{2}{c|}{MLP} & \multicolumn{2}{c|}{CNN} & \multicolumn{2}{c|}{AE} \\
\cline{2-7}
 & Original & Calibrated & Original & Calibrated & Original & Calibrated \\
\hline
\Sys judge (AT + $\eta$) & 0.278 & \textbf{0.558} & 0.411 & \textbf{0.433} & 0.315 & \textbf{0.609} \\
$\eta$ check only & 0.278 & 0.460 & 0.411 & 0.420 & 0.315 & 0.535 \\
AT only & 0.000 & 0.404 & 0.000 & 0.108 & 0.012 & 0.489 \\
\hline
One-Class SVM + $\eta$ & 0.445 & 0.511 & 0.415 & 0.417 & 0.559 & 0.606 \\
One-Class SVM & 0.345 & 0.328 & 0.342 & 0.157 & 0.559 & 0.510 \\
Isolation Forest + $\eta$ & 0.285 & 0.507 & 0.406 & 0.415 & 0.523 & 0.597 \\
Isolation Forest & 0.122 & 0.199 & 0.035 & 0.155 & 0.436 & 0.417 \\
\hline
\end{tabular}%
}
\end{table}

After calibration the judge is the best snapshot detector on all three architectures (Table~\ref{tbl:Calibration}), with F1 0.23, 0.28, and 0.10 above the best baseline alone. Most of that advantage comes from the $\eta$ check: given the same $\eta$ check, the baselines come within 0.05, 0.02, and 0.00. The AT's own contribution is clearest on the MLP, where it catches label poisoning. The calibrated settings are $(\theta, \rho) = (0.08, 0.65)$, $(0.165, 0.40)$, and $(0.12, 0.60)$ for the MLP, CNN, and AE; $\theta$ falls far below 0.3 because honest runs already have 2--4\% of their steps above $\tau$. The 5\% limit does not fully carry over to a new judge: on the unseen seed, honest runs are rejected at 5.2\%, 6.7\%, and 9.6\% on average, mostly by one judge per architecture whose single reference run was atypical. With the spot-check present, calibration does not improve the full decision, so we report the full decision at the original settings.

\subsection{Cost}\label{sec:performance_analysis}

We measure each (architecture, size) over 3 seeds, with runtime on one NVIDIA H100 NVL GPU and peak memory as the peak resident set size on CPU, each stage in its own process. Checkpoint storage is measured from a real checkpoint of an honest run and multiplied by the number of checkpoints each scheme keeps. Verification, which includes the reference run, fitting the judge, calibrating $\delta$, and scoring and spot-checking one submission, costs 1.7--2.3 times one retraining at 300k (Table~\ref{tbl:Cost}) and 1.5--2.3 times across all sizes. \Sys is therefore not cheaper than retraining: its reference run is itself a full-length training run. What it adds is a verdict, which a retraining alone cannot give (\S\ref{sec:recompute}). Beyond the reference run and the judge, checking a submission costs 0.32--1.40\,s of spot-checking plus one pass of the judge. Peak memory stays between 1.3 and 1.8\,GB at every size. Keeping every checkpoint until verification, as in Proof-of-Learning~\cite{jia2021proof}, grows linearly with model size and reaches 3.8\,GB for the 500k MLP; committing one epoch at a time keeps 5--17 times less, and the verifying scientist receives only the sampled segments (16--77\,MB).

\begin{table}[!t]
\centering
\small
\renewcommand{\arraystretch}{1.1}
\setlength{\tabcolsep}{3pt}
\caption{Cost at 300k parameters, means over 3 seeds. Spot-check: re-executing one submission's 10 segments. Storage: every checkpoint kept until verification, against the peak when committing one epoch at a time.}
\label{tbl:Cost}
\resizebox{\ifdim\width>\linewidth\linewidth\else\width\fi}{!}{%
\begin{tabular}{|l|c|c|c|c|c|}
\hline
\multirow{2}{*}{Arch.} & \multicolumn{3}{c|}{Time (s)} & \multicolumn{2}{c|}{Storage (MB)} \\
\cline{2-6}
 & Retrain & Verify & Spot-check & All ckpts & Per epoch \\
\hline
MLP & 24.8 & 51.6 & 0.38 & 2{,}312 & 140 \\
CNN & 7.0 & 12.0 & 0.88 & 292 & 60 \\
AE & 6.2 & 14.6 & 0.34 & 697 & 74 \\
\hline
\end{tabular}%
}
\end{table}

\section{What Each Layer Can Detect}\label{sec:analysis}

\subsection{The Effective-Signal Check Under Gradient Tampering}\label{sec:analysis_eta}

The effective signal $\eta$ measures how much of a step's gradient points in a
consistent direction rather than cancelling out. Write the gradient entries as
$g_1, \dots, g_n$, with mean $\bar g$ and mean square $q = \frac{1}{n}\sum_i g_i^2$;
then $\eta = \bar g^2 / q$.

Sign flipping attacks exactly this quantity. Flipping an entry's sign does not
change its size, so $q$ stays the same, but it pulls the mean toward zero: if each
entry is flipped with probability $s$, the mean shrinks on average to $(1-2s)\bar g$.
Because $\eta$ depends on the square of the mean,
\begin{equation}
\label{eq:eta_flip}
\mathbb{E}[\eta'] \approx (1-2s)^2\,\eta .
\end{equation}
(The exact expression adds $4s(1-s)/n$, which is negligible for models with
$n \ge 10^5$ parameters.) Flipping 10\% of entries therefore cuts the effective
signal to 64\% of its honest value, flipping 30\% cuts it to 16\%, and flipping
50\% removes it entirely.

\begin{figure}[!t]
\centering
\includegraphics[width=\linewidth]{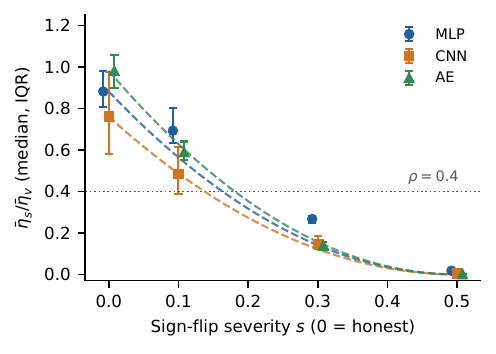}
\caption{How sign flipping lowers a run's effective signal relative to the
verifier's. Points show the median and interquartile range over 45 runs
($s = 0$: the 135 honest runs). Dashed: the prediction, $(1-2s)^2$ times each
architecture's honest median. Runs below the dotted line ($\rho = 0.4$) are rejected.}
\label{fig:eta}
\end{figure}

The $\eta$ check rejects a run whose mean effective signal falls below $\rho$ times
the verifier's. Honest runs sit slightly below the verifier, at a ratio $r_h$, so a
sign-flipped run sits at about $(1-2s)^2 r_h$ and is rejected once $s$ exceeds
\begin{equation}
\label{eq:eta_cut}
s^\ast = \tfrac{1}{2}\left(1 - \sqrt{\rho / r_h}\right).
\end{equation}
With $\rho = 0.4$, this threshold falls between 0.14 and 0.18 depending on the
architecture, and the experiments agree (Figure~\ref{fig:eta}). Every sign flip at
$s = 0.3$ or $0.5$ is rejected. At $s = 0.1$, below the threshold, the check catches
none on the MLP and AE and 29\% on the CNN. The measured ratios track the prediction
closely; the largest gap is the MLP at $s = 0.3$ (0.27 measured against 0.14
predicted), still well below $\rho$.

The $\eta$ check is therefore close to a purpose-built sign-flip detector:
Equation~\eqref{eq:eta_cut} lets a verifier set $\rho$ to catch the weakest sign
flip they care about. It cannot catch fabricated runs, though. Their statistics are
copied from genuine training, so their ratio stays near the honest level.

\begin{figure}[!t]
\centering
\includegraphics[width=\linewidth]{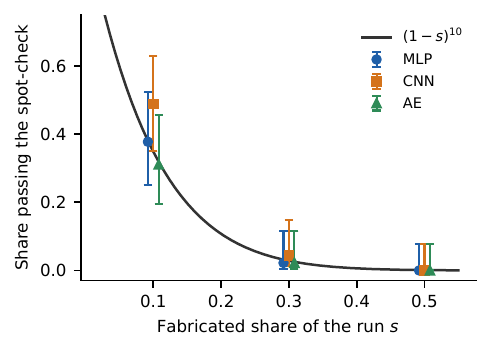}
\caption{Share of fabricated runs passing the spot-check with $k=10$ segments (45 runs per point, 95\% Wilson intervals), against the bound $(1-s)^{10}$.}
\label{fig:spot}
\end{figure}

\subsection{Fabricated Runs Are Invisible to Snapshot Statistics}\label{sec:analysis_fabrication}

A fabricated run's tail is built from the scientist's own genuine statistics, perturbed by noise of the same size as natural step-to-step variation, so any detector that sees only the submitted statistics sees an honest-looking run. No detector separates fabricated runs from honest ones: for the AT, the Isolation Forest, and the One-Class SVM, a randomly chosen fabricated run scores as more anomalous than a randomly chosen honest run with probability at most 0.56 on every architecture, close to the 0.5 of a coin flip, so no choice of threshold can tell them apart. The same detectors do separate other attacks; for example, the AT scores a label-poisoned run above an honest one with probability 0.98 on the MLP and 1.00 on the AE. No calibration of these snapshot detectors can catch fabrication, which is what motivates the spot-check.

\subsection{Soundness of the Spot-Check}\label{sec:analysis_spot}

In a fabricated run of severity $s$, a share $s$ of the segments was never computed. Our forger commits the same checkpoint as both the start and the end of each such segment, so its committed weights never move, while re-running it does move them. Its error is therefore $e_i = 1$: the committed end is off by the segment's entire movement. This exceeds $\delta$ for every judge we trained ($\delta \le 0.84$), so any fabricated segment that is drawn fails. The verifier draws $k$ distinct segments uniformly at random, using randomness the scientist cannot predict, so the run passes only if every drawn segment is one that was actually computed:
\begin{equation}
\label{eq:spot}
\Pr[\text{pass}] \le (1-s)^k,
\end{equation}
because each draw lands on a computed segment with probability at most $1-s$. Drawing without replacement from the run's $N/\ell$ segments makes the exact probability slightly lower. With $k = 10$, the bound predicts that at most 35\%, 3\%, and 0.1\% of fabricated runs pass at $s = 0.1$, 0.3, and 0.5. We measure 31--49\%, 2--4\%, and 0\% (Figure~\ref{fig:spot}). The CNN at $s = 0.1$ is well above the bound, with 22 of 45 runs passing (95\% interval 35--63\%). Because the draws are seeded, we can trace each pass. In 21 of the 22, none of the 10 draws landed on a fabricated segment, which happened in 47\% of runs against an exact probability of 30\% for the CNN's 83 segments. The 22nd run drew the one segment that straddles the point where training stopped. Its first 35 of 50 steps were genuinely computed, so its error was 0.58 rather than 1, below that judge's $\delta = 0.65$. A partly fabricated segment can therefore pass when $\delta$ is loose, which Equation~\eqref{eq:spot} does not account for. Sign-flipped and poisoned runs alter every step, so every drawn segment fails, and none of them passed.  Equation~\eqref{eq:spot} also lets the verifier choose $k$ for a target detection rate. Catching at least 95\% of runs with 10\% fabricated requires $0.9^k \le 0.05$, so $k = \lceil \ln 0.05 / \ln 0.9 \rceil = 29$ segments, or about 1--4\,s of re-execution per submission at our measured cost per segment.

\subsection{Limitations}\label{sec:limitations}

Our evaluation has three seeds per configuration, so calibrated results are measured on held-out seeds, and the CNN and AE margins over the baselines are within seed-to-seed variation. The experiments draw all $k$ spot-check segments when each run starts rather than from ledger randomness, and the ledger commitments themselves are not implemented. This is equivalent for the attacks we test, which do not depend on which segments are drawn, but not for a scientist who watches the draws: if they are revealed one epoch at a time, then once all $k$ have appeared the scientist knows the rest of the run will not be checked and can fabricate it without risk. Equation~\eqref{eq:spot} holds only if the scientist cannot see the draws before committing the whole run. The spot-check also bounds how much of a run can be faked, not whether any single segment was: a scientist who tampers with one segment, for example to plant a backdoor, is caught only if that segment is drawn, with probability $k\ell/N$, which is 1.5--12\% for our runs. In addition, the protocol as described does not require the committed segments to form one continuous run ending at the submitted model, so a scientist could commit segments from one run and submit the model of another; checking that each segment's end weights match the next segment's start, and that the last match the submitted model, needs only the committed hashes, and we leave it to a full implementation. On the CNN, GPU convolutions are not bit-reproducible, so $\delta$ ranges from 0.05 to 0.84; this causes the CNN's 3\% spot-check false positives and leaves less margin against tampering that changes a segment only slightly. The attacks are not adaptive: we test one fabrication strategy and do not evaluate forgers who construct logs designed to pass re-execution~\cite{zhang2022adversarial, fang2023proof}. Our models have at most 500k parameters, and each architecture is paired with a single dataset, so we cannot separate the effect of the architecture from that of the data. We test unauthorized data only as label poisoning; training on data outside the blueprint should also fail re-execution, since the verifier re-runs each drawn segment on the blueprint's data, but we have not tested it directly.

\section{Conclusion and Future Work}\label{sec:conclusion}

We introduced \Sys, which verifies deep learning training within a scientific workflow by letting the scientist explore off the ledger and post a blueprint for each candidate and, instead of requiring exact replication, judging the resulting run in two layers: a statistical judge built from the verifying scientist's own reference run, and a spot-check that re-executes randomly drawn segments from checkpoints committed to the ledger one epoch at a time. On three architectures at 100k--500k parameters and three attacks, \Sys reaches F1 of 0.906--0.976 while rejecting at most 4.4\% of honest runs. The effective-signal statistic falls as $(1-2s)^2$ under sign flipping, so statistics catch strong enough sign flips; fabricated runs are invisible to every snapshot detector we test, and the spot-check catches them with probability at least $1-(1-s)^k$.

Three directions follow. Calibrating each judge on several reference runs should make its false-positive rate hold for new judges. Implementing the per-epoch Merkle commitments on a running SciLedger deployment, together with a segment-selection rule whose draws in one epoch reveal nothing about later epochs, such as checking each segment independently with probability $k\ell/N$, would make the spot-check's guarantee end to end, and recomputing the snapshot statistics of each re-executed segment and comparing them with the submitted ones would tie the statistical judge to the same committed computation. Finally, adaptive adversaries, including forgers that target re-execution~\cite{zhang2022adversarial, fang2023proof}, and a dishonest or colluding verifying scientist, which a reputation-based protocol such as PRISM~\cite{miller_prism_2023} could address, remain open.

\section*{Disclosure on the Use of AI Tools}
Claude (Anthropic), including Claude Sonnet 5 and Claude Opus 5.5, was used in a supporting capacity during the preparation of this work. Specifically, it was used to assist with debugging and reviewing portions of the experimental codebase for implementing the \Sys judge model, spot-check, and evaluation pipeline; parsing, organizing, and statistically analyzing the raw experimental outputs used in \S\ref{sec:experimental_evaluation} and \S\ref{sec:analysis}; generating tables and figures from those outputs; and drafting and organizing portions of the manuscript from author-provided notes, code, and results. The authors independently reviewed, verified, and edited all AI-assisted outputs and take full responsibility for the accuracy, originality, integrity, and content of the manuscript.

\bibliographystyle{IEEEtran}
\bibliography{Citations}

\end{document}